\documentclass[usenatbib]{mnras}

\usepackage{amssymb,amsmath}
\usepackage[T1]{fontenc}
\usepackage{bm}
\usepackage{graphicx}

\title[Dark matter oscillations]{Oscillations of dark matter halos in galaxies and their effects on motion of stars}

\author[Flambaum \& Samsonov]{V. V. Flambaum,
I. B. Samsonov \thanks{E-mail: v.flambaum@unsw.edu.au, igor.samsonov@unsw.edu.au}\\
School of Physics, University of New South Wales, Sydney 2052, Australia
}

\pubyear{\the\year{}}

\begin{document}
\label{firstpage}
\pagerange{\pageref{firstpage}--\pageref{lastpage}}
\maketitle

\begin{abstract}
Matter and dark matter in galaxies represent two main components linked by the gravitational interaction. Collisions of galaxies may create an offset between the centers of mass of these  components. Ignoring internal dynamics of particles in the dark matter halo and Keplerian rotations of matter in the galaxy, we focus on possible relative oscillations of the matter in the dark matter halo. This two-fluid model is somewhat similar to the ``giant dipole resonances'' in nuclei. We estimate possible amplitude and frequency of such oscillations assuming that the offset of the centers of mass is small as compared with the size of the galaxy. Such oscillations, if exist, should manifest themselves in anomalies of velocities of stars in the galaxy, such as the density waves and runaway stars which have orbit periods in resonance with oscillations.
\end{abstract}


\section{Introduction}

An instructive analogy to galactic-scale dark matter -- ordinary matter dynamics can be drawn from nuclear physics. In atomic nuclei, the phenomenon of the \emph{giant dipole resonance} is successfully explained within the liquid-drop model by treating the proton and neutron densities as two overlapping but distinct fluids whose centers of mass can oscillate relative to one another under external perturbations. A similar two-fluid picture may be invoked for galaxies, where the matter and dark matter components can, in principle, respond differently to dynamical or environmental influences. A striking demonstration of such differential behavior is provided by the famous \emph{Bullet Cluster} (1E~0657--558), in which the collision of two galaxy clusters has separated the hot intracluster gas (traced by X-ray emission) from the dark matter halos, whose gravitational potential is mapped by weak lensing \citep{Clowe:2006eq}. Following the collision, the hot gaseous component is observed to lag behind due to collisional interactions, while the dark matter behaves as a nearly collisionless fluid and continues almost unimpeded. Another striking observation of matter -- dark matter interaction was provided by \cite{Jee_2007,Jee_2010} for the galaxy cluster Cl 0024+17 where a ring-like dark matter structure was found. These observations are regarded as direct empirical evidence for dark matter. 

In the present work, we extend this two-fluid analogy to the internal dynamics of a model galaxy by considering the matter and dark matter halo as two gravitationally bound continua whose centers of mass may experience relative motion. Our aim is to estimate and characterize the possible oscillations of these two components as a whole, neglecting the internal density fluctuations and the Keplerian motion of individual stars. Of special interest is whether such large-scale relative oscillations could manifest observationally as systematic anomalies in the stellar velocity field of the Milky Way and other spiral galaxies, such as those discovered by \cite{Poggio25}. This framework naturally motivates the subsequent discussion of theoretical predictions and observational constraints on the displacement between the centers of the visible and dark components in the Milky Way.

Numerical simulations and observational data indicate that the center of a galactic dark matter (DM) halo does not necessarily coincide with the center of the visible baryonic disc. High-resolution cosmological simulations of Milky Way-type galaxies suggest that the DM density peak may be displaced by several hundred parsecs from the baryonic center. In particular, \citet{Kuhlen2013} found an offset of approximately $300$--$400~\mathrm{pc}$ between the halo and stellar disc centers in the Eris hydrodynamic simulation, a result attributed to the interaction between the galactic bar and the inner DM halo. Further studies in large-volume cosmological runs such as the EAGLE suite by \cite{Schaller2015} show that most galaxies exhibit smaller offsets, typically below the gravitational softening length of $\sim700~\mathrm{pc}$, although transient displacements of a few hundred parsecs may occur following mergers or tidal perturbations.

Observational constraints on such offsets are more challenging. Strong-lensing analysis of a cluster member galaxy in Abell~3827 revealed a projected separation of $1$--$6~\mathrm{kpc}$ between the stellar and DM mass peaks \citep{Massey2015}, interpreted as potential evidence for dark-matter self-interaction, although later reanalyses revised the magnitude and significance of this result. Within the Milky Way, the spatial distribution of $\gamma$-ray emission has been used to probe the DM density center. The peak of the previously claimed $130~\mathrm{GeV}$ line was offset by about $200~\mathrm{pc}$ from Sgr~A$^\ast$ \citep{Su:2012ft}, consistent with expectations from simulations done by \citet{Kuhlen2013}. Baryon sloshing effects were uncovered upon the analysis of \emph{Gaia} space observatory data by \cite{Belokurov-22} and in recent numeric simulations presented by \cite{Bland-Hawthorn_2025}. Analytical modeling by \citet{Prasad2017} demonstrated that a modest halo offset of $\sim350~\mathrm{pc}$ can generate strong $m=1$ lopsidedness in the inner disc, producing kinematic and morphological asymmetries similar to those observed in the central few kiloparsecs of the Milky Way. Taken together, both numerical and observational studies suggest that a displacement of order $200$--$400~\mathrm{pc}$ between the DM and baryonic centers is plausible for Milky Way-like galaxies, while larger separations are unlikely.

Despite the conceptual analogy, the oscillatory behavior of the matter and dark matter components in galaxies differs from the giant dipole resonance in atomic nuclei. In the nuclear case, the restoring force arises from the strong interaction between positively charged protons and electrically neutral neutrons, giving rise to an electric dipole mode that couples efficiently to the dipole component of the external electromagnetic field.  The resulting nuclear giant dipole resonance therefore represents a collective oscillation of charge. There are also collective  resonances associated with higher multipoles. They  can decay rapidly through  excitation of nuclear internal degrees of freedom, such as single-particle and  phonon excitations, with the nuclear giant resonances width comparable to the resonances excitation energy. The nuclear giant resonance is actually an envelope covering a very dense spectrum of narrow resonances. 

The mechanism of the matter -- dark matter oscillations damping with the energy transfer to the motion of gas, stars and dark matter particles may also work. Gravity is always attractive and lacks the corresponding dipole moment; the lowest radiative multipole of the gravitational field is the quadrupole. Consequently, relative dipole oscillations between DM and matter distributions cannot be excited by the dipole gravitational field of a distant passing galaxy, higher multipoles are needed. The dipole deformations of the hot gas component may be most efficiently excited by a close encounter of two galaxies, like that in the Bullet Cluster \citep{Clowe:2006eq} or in Cl 0024+17 \citep{Jee_2007,Jee_2010}.

The damping of matter -- dark matter oscillations is governed primarily by gravitational interactions between the two components, manifesting as dynamical friction introduced originally by \cite{Chandrasekhar43}. This relatively slow dissipation implies that the oscillations, once excited, may persist over cosmological timescales, potentially leaving observable kinematic signatures in the velocity field of stars and gas within galactic discs. Similar effects were observed recently in \emph{Gaia} space observatory data by \cite{Belokurov-22} and in numerical simulations presented by \cite{Bland-Hawthorn_2025} (``baryon sloshing''). In the present paper, we show that the leading dipole mode in the relative dynamics of the matter and dark matter components allows for a simple analytic description which provides reasonable accuracy and is consistent with numerical simulations.

The rest of the paper is organized as follows. In the next section, we consider a two-component model of a galaxy with some density distributions for matter and dark components. Assuming that these components appear out of equilibrium due to some external perturbation, we show that their relative displacement, in the leading-order approximation, is given by damped harmonic oscillations. We give analytic expressions and some numerical estimates for the frequency of these oscillations and the damping factor assuming that the latter is dominated by the dynamical friction. We show that for typical galactic mass parameters close to the ones of the Milky Way galaxy, the oscillation frequency is comparable with the star rotation periods while damping is not too strong so that these oscillations may last on the cosmological timescale. In Sec.~\ref{SecTrajectories}, we present one of the possible observable effects of the dark matter halo oscillations: perturbations of star orbits due to the oscillations of the dark matter halo relative to the baryonic matter. We show that some star orbits may appear in resonance with the dark matter halo oscillations, leading to strong perturbations of the orbits and to observable deformations of the galactic disc. In the phase space, we show that these effects manifest themselves in the asymmetry of the $(z,v_z)$ subspace (assuming that the perturbation is applied in the $z$ direction) which becomes more apparent for the star trajectories with periods in resonance with DM halo oscillation. In Sec.~\ref{SecGas}, we consider oscillations of the hot gas halo relative to the matter and dark matter components. We show that the oscillations of the hot gas component exhibit only minor effects in the galaxy dynamics, as compared with the effects of oscillations of the dark matter halo relative to the matter. In the last section, we discuss the results and open questions.


\section{Relative oscillations of two misaligned spherical galactic components}
\label{sec:spherical_oscillations}

We begin with a simplified model consisting of two spherically symmetric mass distributions representing matter (stars and cold gas clouds) and dark matter components of a galaxy. Each component is treated as a self-gravitating fluid with equilibrium density profiles $\rho_m(r)$ for matter and $\rho_d(r)$ for dark matter, respectively. In equilibrium, the two components are concentric, sharing a common center of mass. We now consider a small relative displacement $\boldsymbol{\delta}$ between their centers, such that $|\boldsymbol{\delta}| \ll R_m, R_d$, where $R_m$ and $R_d$ denote the characteristic radii of the matter and dark matter distributions. The aim of this section is to derive the restoring force and oscillation frequency of this relative displacement mode.

In the leading-order approximation, we ignore deformations of density profiles of the matter and dark matter components and disregard internal dynamics in these components. We focus on their relative dynamics.

\subsection{Equilibrium configuration}

Each component obeys the Poisson equation
\begin{equation}
\nabla^2 \Phi_i = 4\pi G \rho_i(r), 
\qquad i=m,d,
\end{equation}
with the corresponding gravitational potentials $\Phi_b(r)$ and $\Phi_d(r)$. $G$ is the gravitational constant. In equilibrium, the total gravitational potential
\begin{equation}
\Phi_0(r) = \Phi_m(r) + \Phi_d(r)
\end{equation}
is spherically symmetric, and the total system is static. 

For definiteness, we consider the dark matter density either in the pseudo-isothermal form
\begin{equation}
\rho_d(r) = \frac{\rho_0}{1+(r/R_c)^2}\,,
\label{pseudo-isothermal}
\end{equation}
or, for comparison, in the Navarro--Frenk--White (NFW) form
\begin{equation}
\rho_d(r) = \frac{\rho_s}{(r/R_s)(1+r/R_s)^2}\,.
\label{NFW}
\end{equation}
The matter component (stars plus cold gas clouds) may be approximated by the Plummer profile,
\begin{equation}
\rho_m(r) = \frac{3M_b}{4\pi R_m^3}\!\left(1+\frac{r^2}{R_m^2}\right)^{-5/2}\,.
\label{Plummer}
\end{equation}
These analytic forms reproduce the essential structure of the matter and the extended dark halo in the galaxy.

\subsection{Perturbed configuration and potential expansion}

Let the matter component be displaced by $\boldsymbol{\delta}$ relative to the dark halo. 
The gravitational potential of the dark component at a point $\boldsymbol{r}$ in the matter frame is
\begin{equation}
\Phi_d(|\boldsymbol{r}+\boldsymbol{\delta}|) = \Phi_d(r) + \boldsymbol{\delta}\!\cdot\!\nabla \Phi_d(r)
+ \mathcal{O}(\delta^2)\,.
\end{equation}
The linear term represents a perturbation potential acting on the baryonic distribution. Since $\Phi_d$ is spherically symmetric, $\nabla \Phi_d = (d\Phi_d/dr)\,\hat{\boldsymbol{r}}$, and, representing $\boldsymbol{\delta}\cdot\hat{\boldsymbol{r}} = \delta \cos\theta$, the perturbation can be written as
\begin{equation}
\Phi_{\text{pert}} := 
\boldsymbol{\delta}\!\cdot\!\nabla \Phi_d(r)
=
\delta\,\frac{d\Phi_d}{dr}\cos\theta\,,
\label{PhiPert}
\end{equation}
which corresponds to a dipole ($l=1$) distortion of the potential. 
An equivalent expansion for the dark matter potential in the frame of the displaced halo was obtained by \cite{Prasad2017}, who showed that such a displacement produces an azimuthal $m=1$ perturbation in the disk plane. 
In the present spherical model, the perturbation retains the same dipolar structure.

\subsection{Restoring force and oscillation frequency}

The gravitational force acting on the center of mass of the matter component due to the perturbation (\ref{PhiPert}) is obtained by integrating the perturbed potential over its density:
\begin{equation}
\boldsymbol{F} 
= -\int \rho_b(\boldsymbol{r}) 
\nabla \Phi_d(|\boldsymbol{r}+\boldsymbol{\delta}|)\, d^3r\,.
\label{force}
\end{equation}
Expanding to first order in $\boldsymbol{\delta}$, the net restoring force is
\begin{equation}
\boldsymbol{F} = -k\, \boldsymbol{\delta}\,, 
\end{equation}
where
\begin{align}
k =& \frac{4\pi}{3} \int_0^{\infty} r^2 \rho_m(r)\,\nabla^2\Phi_d(r)\,dr
\nonumber \\ =& \frac{16\pi^2 G}{3}\! \int_0^{\infty} r^2 \rho_m(r)\,\rho_d(r)\,dr\,.
\label{k-integral}
\end{align}
Similarly, the reaction force acting on the dark component due to the baryonic potential yields
$k\, \boldsymbol{\delta}$.

If the matter component is much smaller in size and mass than the dark matter component, $M_m\ll M_d$, $R_m\ll R_d$, then the baryonic component may be roughly considered as a point-like mass, $\rho_m\approx M_m \delta^3(\boldsymbol{r})$, and the expression (\ref{k-integral}) is simplified to $k = \frac{4\pi}{3}G M_m \rho_d(0)$.

The equations of motion for the centers of mass of each component, $\boldsymbol{r}_m$ and $\boldsymbol{r}_d$, are
\begin{align}
M_m \ddot{\boldsymbol{r}}_m =& -k(\boldsymbol{r}_m-\boldsymbol{r}_d)\,, 
\\
M_d \ddot{\boldsymbol{r}}_d =& -k(\boldsymbol{r}_d-\boldsymbol{r}_m)\,.
\end{align}
Eliminating the common center-of-mass motion, one finds the equation for the relative displacement 
$\boldsymbol{\delta}=\boldsymbol{r}_m-\boldsymbol{r}_d$,
\begin{equation}
\ddot{\boldsymbol{\delta}} + \omega_0^2\, \boldsymbol{\delta} = 0\,,
\label{delta-equation}
\end{equation}
where the oscillation frequency is
\begin{equation}
\omega_0^2 = \frac{k}{\mu}\,,\qquad
\mu = \frac{M_m M_d}{M_m+M_d}\,.
\end{equation}
In the limit $M_m\ll M_d$, we have $\mu\approx M_m$, and the oscillation frequency becomes $\omega_0^2 \approx \frac{4\pi}{3}G\rho_d(0)$.

More generally, equation (\ref{delta-equation}) may involve the damping term due to dynamical friction in the baryonic component of the galaxy,
\begin{equation}
    \ddot{\boldsymbol{\delta}} + 2\gamma\omega_0 \dot{\boldsymbol{\delta}} + \omega_0^2\boldsymbol{\delta} = 0\,.
    \label{harm-oscillator}
\end{equation}
The damping parameter $\gamma$ is estimated in Sec.~\ref{SecFriction} below.

In the above derivation, the spherical distribution of matter and dark matter in the galaxy was assumed. For non-spherically-symmetric galaxies, equation (\ref{force}) is modified as 
$F_i = -k_{ij}\delta_j$, with $k_{ij} = \int \rho_b \partial_i \partial_j\Phi d^3r$. To avoid this complication, we assume that the mass and size parameters of the dark matter halo are much larger than the corresponding parameters of the matter component, $M_m\ll M_d$, $R_m\ll R_d$, and in the leading-order approximation it is possible to model the visible matter as a point-like mass moving in the potential created by the dark matter halo. For small deviation of this mass from the equilibrium position, the relative dynamics is described by Eq.~(\ref{harm-oscillator}).


\subsection{Numerical estimate of oscillation period}
\label{Estimates}

For an order-of-magnitude estimate, in this section we consider a galaxy with macroscopic parameters close to the ones of the Milky Way galaxy. We assume that the matter mass distribution is given by the Plummer formula (\ref{Plummer}) with $M_m = 6\times 10^{10}M_\odot$ and $R_m = 10$\,kpc. For the dark matter distribution, we assume the cored profile (\ref{pseudo-isothermal}) with the core parameter $R_c = 2$\,kpc and the central density $\rho_0 = 0.25\,M_\odot\mathrm{pc}^{-3}$. These parameters correspond to the total dark matter mass $M_d \simeq 10^{12}\,M_\odot$ enclosed within the sphere of radius 100 kpc, and the distant star rotation velocities of about $220$\,km\,s$^{-1}$.

For these density profiles, the integral (\ref{k-integral}) may be calculated analytically,
\begin{align}
    k= &\frac{2\pi M_m G \rho_0 \xi^3 }{3(1-\xi^2)^{5/2}}\big( 6\arcsin\xi 
    \nonumber\\&
    +2(\xi+2\xi^{-1})\sqrt{1-\xi^2} -3\pi \big),
\end{align}
with $\xi = R_c/R_m=0.2$. As a result, for this density profile we find the oscillation period
\begin{equation}
    T=\frac{2\pi}{\omega_0} \simeq 390\,\mbox{Myr}.\quad (\mbox{cored profile})
    \label{Tcored}
\end{equation}

In a similar way, we consider the NFW dark matter profile (\ref{NFW}) with the following values of the parameters: $R_s = 60$\,kpc, $\rho_s=10^{-3}\,M_\odot\mathrm{pc}^{-3}$. Evaluating the integral (\ref{k-integral}) we find $k\simeq 1.2\times 10^{-5}\,M_\odot\mbox{yr}^{-2}$, and the corresponding oscillation period is close to the result (\ref{Tcored}):
\begin{equation}
    T \simeq 440\,\mbox{Myr}. \qquad (\mbox{NFW profile})
    \label{TNFW}
\end{equation}
These values are comparable to the Galactic rotation periods. They are weakly sensitive to changes of $R_m$ and the halo parameters, because the integral (\ref{k-integral}) is dominated by the spatial overlap of the baryonic core and the inner halo.

The authors \cite{Kuhlen2013,Schaller2015,Massey2015,Su:2012ft,Prasad2017} argued that offsets of centers of mass in dark matter and baryonic matter components in typical galaxies are likely of order 200-400 pc. Therefore, for our estimates we assume displacements of the centers of mass oscillates 
\begin{equation}
\delta(t) = \delta_0 \sin(\omega_0 t)\,,
\label{perturbation}
\end{equation}
with amplitude 
\begin{equation}
\delta_0\simeq 400\,\text{pc}.
\label{delta0}
\end{equation}
The corresponding relative velocity between these two components evolves as
\begin{equation}
v_{\rm rel}(t) = \omega_0\,\delta_0\cos(\omega_0 t)\,.
\end{equation}
The root-mean-squared velocity is
\begin{equation}
    \bar v:= \langle v_\mathrm{rel}^2 \rangle^{1/2} = \frac1{\sqrt2}\omega_0\delta_0\,.
\end{equation}
For the characteristic oscillation frequency $\omega_0 = 0.016\,\mathrm{Myr}^{-1}$, we have 
\begin{equation}
\bar v \simeq 4\,\mathrm{km\,s}^{-1}\,.
\end{equation}
This speed is much smaller than the velocity of the Sun in the galactocentric frame, $v_\odot \approx 250$~km\,s$^{-1}$. Therefore, such oscillations, if exist, are hardly observable in laboratory experiments on the Earth. However, if the initial displacement of the centers of the baryonic and dark matter is not small, this velocity can represent a significant effect of dark matter wind relative to the Earth.

\subsection{Estimate of the damping parameter}
\label{SecFriction}

The dynamical friction may be understood as the gravitational drag force experienced by the matter component with mass $M_m$ in the galaxy when it moves with velocity $\boldsymbol{v}=\dot{\boldsymbol{\delta}}$ through the gas of dark matter particles with density $\rho_d$. This force may be roughly estimated using formula by \cite{Chandrasekhar43}:
\begin{equation}
    \boldsymbol{F}_\mathrm{df} = -4\pi G^2 M_m^2 \rho_d \frac{\log\Lambda}{v^3}\left[ \mathrm{erf}(X) - X\mathrm{erf}'(X) \right]\boldsymbol{v}\,,
    \label{Fdf}
\end{equation}
where $X := \frac{v}{\sqrt2 \sigma}$, $\sigma$ is the dark matter particles velocity dispersion, and $\ln\Lambda$ is the Coulomb logarithm which accounts for the range of gravitational encounters.

Since we are considering small oscillations of the baryonic and dark components in the galaxy near the common center of mass, the corresponding velocities $v$ are much smaller than the dark matter particles velocity dispersion, $v\ll \sigma$. In this regime, $X\ll1$, and $\mathrm{erf}(X) - X \mathrm{erf}'(X)\approx \frac{4X^3}{3\sqrt\pi}=\frac{\sqrt2 v^3}{3\sqrt{\pi}\sigma^3}$. Hence, the dynamical friction (\ref{Fdf}) is simplified:
\begin{equation}
    \boldsymbol{F}_\mathrm{df} \approx -\frac{4\sqrt{2\pi} G^2 M_m^2 \rho_d \ln\Lambda}{3\sigma^3}\dot{\boldsymbol{\delta}}\,.
    \label{Fdf-simple}
\end{equation}
Comparing this with Eq.~(\ref{harm-oscillator}), we find the expression for the oscillation damping parameter
\begin{equation}
    \gamma = \frac{2\sqrt2 G^2 M_m \rho_d \ln\Lambda}{3\omega_0 \sigma^3}\,.
    \label{damping}
\end{equation}

For our numerical estimate, we take $M_m = 6\times 10^{10}\,M_\odot$, $\rho_d\simeq 10^{-3}\,M_\odot\mbox{pc}^{-3}$, $\ln\Lambda\simeq 5$, $\omega_0 = 2\pi/T\simeq (\pi/200)\mathrm{Myr}^{-1}$, $\sigma\simeq 200$\,km\,s$^{-1}$. This gives the damping coefficient and the corresponding quality factor:
\begin{equation}
    \gamma\approx 0.05\,,\qquad
    \gamma^{-1} \approx 20\,.
    \label{gamma}
\end{equation}
Thus, it is feasible that, in galaxies, the baryonic component can do $\sim$10 oscillations in the gravitational potential of the dark matter halo before relaxation. These estimates are in agreement with similar estimates and numerical simulations by \cite{Bland-Hawthorn_2025}.

The estimates (\ref{gamma}) may be considered as a lower bound on the damping constant. In reality, this damping may be higher because of the decay of the dipole mode into higher multipoles representing internal dynamics in the matter and dark galactic components. Detailed study of the applicability of the Chandrasekhar’s formula (\ref{Fdf}) to the galaxy mergers was done by \cite{Vasiliev_2022}. Accurate estimates of these effects are beyond the scope of this paper. For realistic galaxies, the value of the damping constant may be taken either from observations or from accurate numerical simulations.


\section{Perturbations of star trajectories by oscillating dark matter halo}
\label{SecTrajectories}

In this section we consider perturbations of trajectories of stars due to oscillations of center of mass of the matter component of the galaxy relative to the dark matter halo center of mass. In disc galaxies, it is of interest to study perturbations of trajectories of stars both orthogonal to the galactic disc ($z$-direction) and along the disc ($R$-direction). Here, for simplicity, we assume that the dark matter halo oscillates in the $z$ direction thus creating perturbations orthogonal to the galactic disc. Another natural assumption is $M_m\ll M_d$, meaning that the motion of the halo in the center of mass frame may be ignored, although this assumption is not important and may be easily relaxed.

\subsection{Configuration space analysis}

In the galactic moving frame, the dark matter center of mass coordinate oscillates by the rule \begin{equation}
\boldsymbol{r}_d(t) = \hat{\boldsymbol{z}} Ae^{-\gamma\omega_0 t}\sin(\omega_0 t)\,,
\label{rd}
\end{equation}
with some amplitude $A\ll R_m$ and angular velocity $\omega_0$. Stars in the galaxy move in the time-dependent gravitational potential
\begin{align}
\Phi(\boldsymbol{r},t) = &\Phi_m(\boldsymbol{r}) + \Phi_d(\boldsymbol{r}-\boldsymbol{r}_d(t))\nonumber\\
\approx&\Phi_m(\boldsymbol{r}) + \Phi_d(\boldsymbol{r}) - \boldsymbol{r}_d(t)\cdot \nabla \Phi_d(\boldsymbol{r})\,.
\end{align}
Trajectories of the stars in this potential are governed by the Newton's equation, 
\begin{equation}
\ddot{\boldsymbol{r}} = -\nabla\Phi(\boldsymbol{r},t) + \ddot{\boldsymbol{r}}_d\,,
\label{NewtonEq}
\end{equation}
with the last term taking into account the acceleration of the galactic frame.

Equation of motion for stars in the oscillating dark matter halo (\ref{NewtonEq}) may be solved numerically. Dynamics of the $z$-component in this equation, however, may be estimated analytically for small deviations from the non-perturbed trajectories. Denoting $\boldsymbol{R}=(x,y)$, we consider the $z$-component of Eq.~(\ref{NewtonEq}),
\begin{equation}
    \ddot z = -\partial_z\Phi_m(\boldsymbol{R},z) - \partial_z\Phi_d(\boldsymbol{R},z-z_d) + \ddot z_d\,,
\end{equation}
and linearize it in $z$ and in $A$,
\begin{equation}
    \ddot z + \omega^2 z = B e^{-\gamma \omega_0 t} \sin (\omega_0 t + \varphi)\,,
\end{equation}
where
\begin{align}
    \omega^2 =& \partial_z^2 \Phi_m(\boldsymbol{R},0) + \partial_z^2 \Phi_d(\boldsymbol{R},0) \,,\\
    B =&A \sqrt{(\partial_z^2 \Phi_d(\boldsymbol{R},0)-(1-\gamma^2)\omega_0^2)^2+ 4\gamma^2\omega_0^4}\,,\\
    \varphi =&-\mathrm{arctan}\frac{2\gamma \omega_0^2}{\partial_z^2 \Phi_d(\boldsymbol{R},0)-(1-\gamma^2)\omega_0^2}\,.
\end{align}
We conclude that the dynamics in the $z$ direction is approximately given by small oscillation with proper frequency $\omega$ and driving force with frequency $\omega_0$. Notice that the latter frequency is fixed while $\omega = \omega(R)$ varies with distance from the center $R$. Therefore, for some star trajectories, there may be resonance amplification of the amplitude of oscillations near
\begin{equation}
    \omega(R) = \omega_0\,. 
    \label{omegaR}
\end{equation}

\begin{figure}
    \centering
    \includegraphics[width=\linewidth]{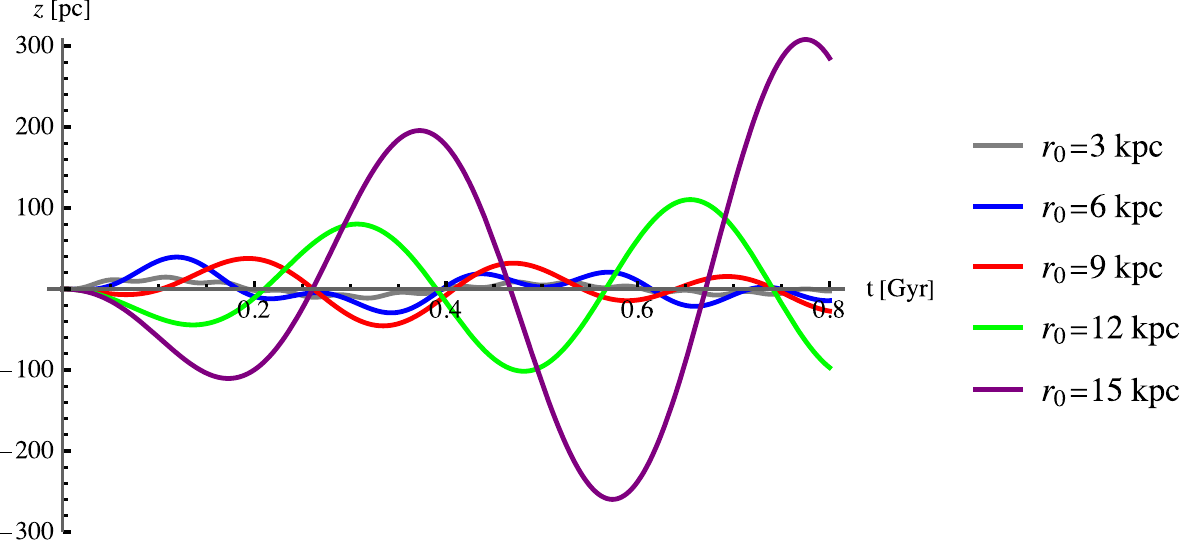}
    \caption{Time evolution of the vertical ($z$) displacement of circular stellar trajectories initially located at the distance $r_0$ from the galactic center. Without perturbation, all stars remain in the plane $z=0$; once the perturbation (\ref{perturbation}) is activated ($t>0$), the trajectories develop oscillatory vertical motion.}
    \label{Fig1}
\end{figure}

\begin{figure}
    \centering
    \includegraphics[width=0.9\linewidth]{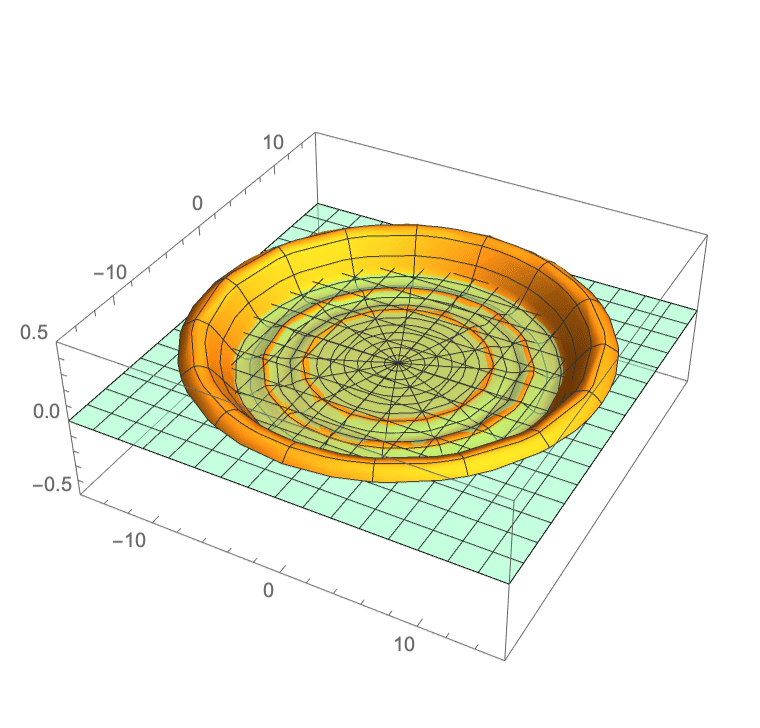}
    \caption{Deformations of the galactic disc due to oscillating dark matter halo in the direction orthogonal to the disc after $0.7$ Gyr since the periodic dark matter oscillation was introduced. Units are kpc.}
    \label{Fig2}
\end{figure}

An illustration of the effect of perturbation of star orbits in the galaxy is given in Fig.~\ref{Fig1}. In this figure, we consider a model galaxy with parameters from Sec.~\ref{Estimates} above, and assume that the periodic dark matter halo oscillation (\ref{rd}) starts at $t=0$ in the $z$-direction. The period of these dark matter oscillations is taken $T = 400$\,Myr, which is consistent with the estimates (\ref{Tcored}) and (\ref{TNFW}). This perturbation sources vertical displacements of trajectories of stars which may be found from numerical solution of the equation (\ref{NewtonEq}). In Fig.~\ref{Fig1}, we present time evolution of the $z$-coordinate of a few trajectories of stars at the distance of 3-15 kpc from the galactic center (GC). For simplicity, we consider deformations of circular orbits in the galactic plain with $z=\dot z =0$ at $t=0$. Typical deformation of a galactic disc due to this perturbation is shown in Fig.~\ref{Fig2}.

\begin{figure}
    \centering
    \includegraphics[width=0.9\linewidth]{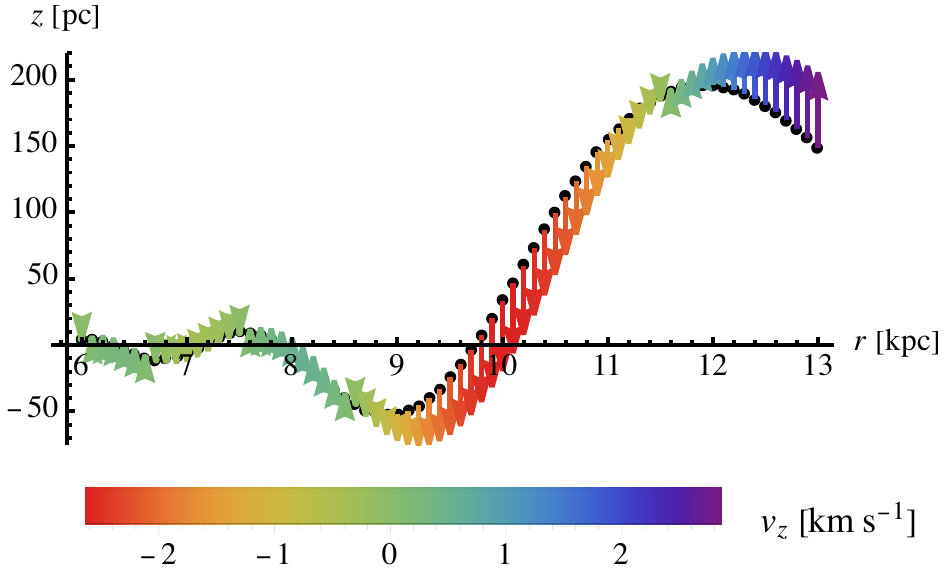}
    \caption{Vertical displacements of stars from the galactic plain due to the oscillating DM halo at the moment $t=0.6$\,Gyr from the beginning of vertical periodic perturbation as in Eq.~(\ref{rd}). The colour-coded arrows represent the component of the star velocity $v_z$.}
    \label{Fig25}
\end{figure}

Fig.~\ref{Fig1} shows that the amplitude of the vertical perturbations of trajectories of stars grows with distance $R$ from GC. This is explained by the fact that the trajectories at the distance $R=14.8$ kpc from GC appear in resonance with the dark matter oscillations. This distance is found from numerical solution of the equation (\ref{omegaR}) with the accepted parameters of the matter and dark matter distributions considered in the present section.

We stress that the analysis in this section serves as a demonstration of the effect of resonant enhancements of $z$-perturbations of orbits of some stars if their orbital periods are close to the oscillation period of the dark matter halo. Analytic approximations and numeric estimates in this section show that this resonance is possible for realistic values of parameters (mass, size and density distribution) of actual galaxies. This effect manifests itself mainly in vertical displacement of stars in disc galaxies. 

In Fig.~\ref{Fig25}, we plotted vertical displacements of stars with nearly circular orbits at the distance $6\,\text{kpc}<r<13\,\text{kpc}$ from GC and at the moment $t=0.6$\,Gyr from the beginning of the vertical perturbation (\ref{rd}). Coloured arrows in this figure represent the vertical components of the velocities of stars. The curve in Fig.~(\ref{Fig25}) has an apparent maximum near $r=12$\,kpc, and the spread of the velocities is $|v_z| \lesssim 3\,\text{km\,s}^{-1}$. Similar wave-like structure for anomalous vertical components of velocities of the stars in the Milky Way galaxy is observed by \cite{Poggio25}, see Fig. 16. The observed values of $|\Delta z|\lesssim 0.2$\,kpc and $|v_z|\lesssim 3$\,km\,s$^{-1}$ are very close to the ones predicted by our model. This similarity suggests that an oscillating dark matter halo could contribute to Great-Wave-like vertical corrugations. Our analysis, however, does not exclude other non-axisymmetric perturbations in the Milky Way disc producing residual velocities and displacement of comparable or larger magnitude.

More generally, it is possible to study oscillations of dark matter halo at arbitrary angle to the galactic disk. The component of the oscillating displacement vector $\boldsymbol{\delta}$ parallel to the disc can source the contributions to the radial components of the velocities of stars in the galactic plain.

\subsection{Observational uncertainties and intrinsic scatter}

The amplitudes found above should be compared with the observational and astrophysical noise floor of the relevant
stellar tracers. For the Milky Way, the most direct comparison is with the young-giant and Cepheid samples of \cite{Poggio25}. The young-giant sample contains about $1.7\times 10^4$ Gaia DR3 stars out to heliocentric distances $d_{\rm helio}\lesssim 6$--$7\,{\rm kpc}$, while the Cepheid sample contains about $3.4\times 10^3$ objects out to $d_{\rm helio}\lesssim 15\,{\rm kpc}$.  For Cepheids the typical distance uncertainty is below $5\%$, although it can reach $\sim 13\%$, while for the young giants about $90\%$ of the sample has relative distance error below $20\%$. Since $z=d\sin b$, these distance uncertainties correspond to vertical position errors of order $10$--$90\,{\rm pc}$ for Cepheids and up to $\sim 40$--$140\,{\rm pc}$ for young giants over the range $|z|\sim 0.2$--$0.7\,{\rm kpc}$.  In practice, however, the individual-star scatter in $z$ is dominated by the intrinsic thickness of the young disc, which \cite{Poggio25} model with a fiducial value of about $150\,{\rm pc}$ and a systematic range $100$--$300\,{\rm pc}$.

The corresponding velocity errors are also tracer-dependent. For bright Gaia DR3 stars with $G<15$, the proper-motion uncertainty $\sigma_\mu\simeq 0.02$--$0.03\,{\rm mas\,yr}^{-1}$ gives a transverse-velocity error $\sigma_{v_\perp}=4.74\,\sigma_\mu d_{\rm helio}$, i.e. about $0.3$, $1.0$, and $2.1\,{\rm km\,s}^{-1}$ at $d_{\rm helio}=2$, $7$, and $15\,{\rm kpc}$, respectively. Gaia DR3 radial-velocity uncertainties range from about $1.3\,{\rm km\,s}^{-1}$ at $G_{\rm RVS}=12$ to $6.4\,{\rm km\,s}^{-1}$ at $G_{\rm RVS}=14$ \citep{Gaia3}.  Thus the formal observational error in $v_z$ is typically of order $0.5$--$2\,{\rm km\,s}^{-1}$ for bright young giants and can reach a few ${\rm km\,s}^{-1}$ for distant Cepheids.

The dominant limitation is not the formal Gaia error but the intrinsic dynamical scatter of the young Galactic disc. A representative single-star noise floor is $\sigma_{v_z}\sim 5$--$10\,{\rm km\,s}^{-1}$ and $\sigma_{v_R}\sim 10$--$20\,{\rm km\,s}^{-1}$, reflecting birth velocities, spiral structure, the warp, satellite perturbations and other non-axisymmetric motions. These random components are reduced in binned maps approximately as $N^{-1/2}$, so coherent features may be detected with uncertainties of order $1$--$3\,{\rm km\,s}^{-1}$ in $v_z$ and $20$--$60\,{\rm pc}$ in $z$ for typical bins containing tens of tracers. Consequently, the perturbation predicted in our toy model, $|\Delta z|\lesssim 0.2\,{\rm kpc}$ and $|v_z|\lesssim 3\,{\rm km\,s}^{-1}$, is not a single-star signature. It is potentially observable only as a coherent large-scale phase-space pattern after subtracting the smooth warp and rotation field.

For external galaxies the situation is less clear. Except for the nearest Local Group systems, individual stellar phase-space coordinates are generally unavailable, and one must rely on integrated-light spectroscopy or discrete tracers. At galactocentric radii $R\gtrsim 12\,{\rm kpc}$, where the signal in our model is largest, the declining surface brightness and tracer density increase the observational uncertainties. A few-${\rm km\,s^{-1}}$ perturbation would therefore be difficult to isolate in a single external galaxy. More promising tests would involve statistical samples of galaxies with similar morphology and interaction history, or controlled cosmological simulations in which the intrinsic scatter can be measured directly.


\begin{figure*}
    \centering
    \setlength{\tabcolsep}{6pt}
    \begin{tabular}{cc|c|c}
         & $R_0=9$\,kpc & $R_0=12$\,kpc & $R_0=15$\,kpc \\
       \rotatebox{90}{perturbation=off}& \includegraphics[width=4cm]{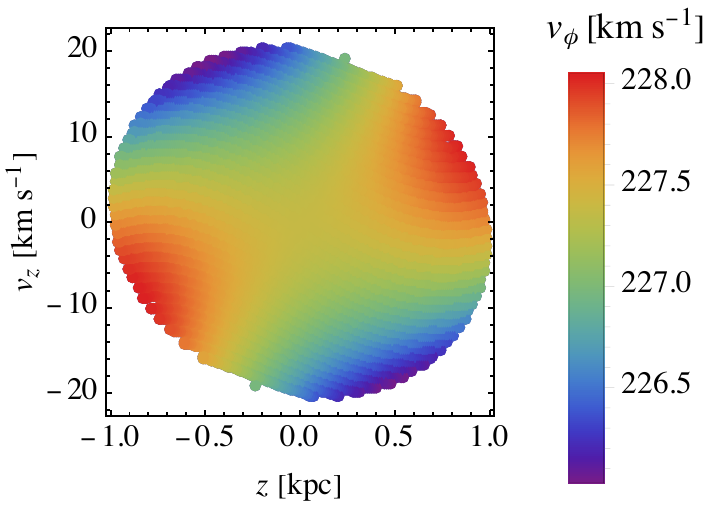} {\hspace{7pt}} & {\hspace{7pt}} \includegraphics[width=4cm]{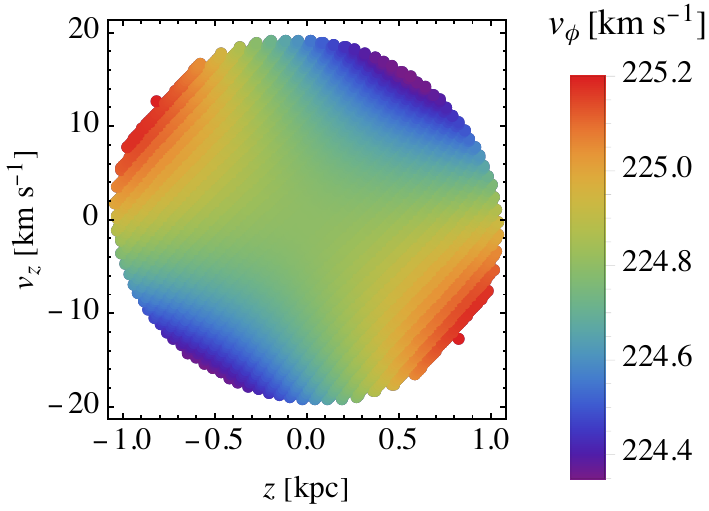} {\hspace{7pt}} & {\hspace{7pt}} \includegraphics[width=4cm]{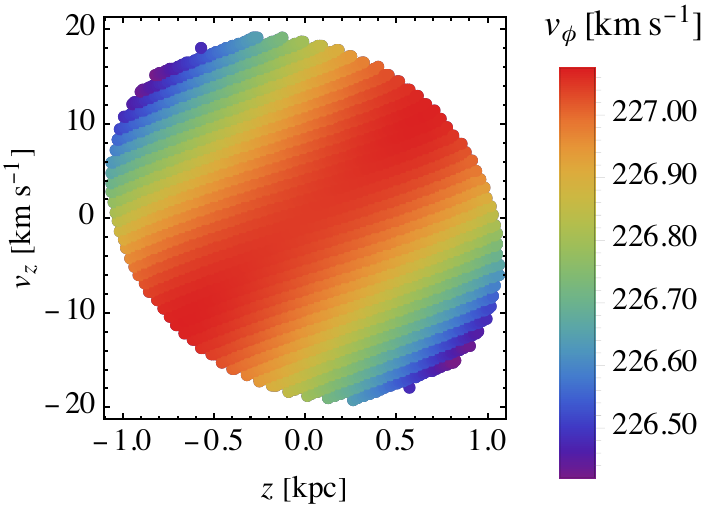} \\
       \rotatebox{90}{perturbation=on}& \includegraphics[width=4cm]{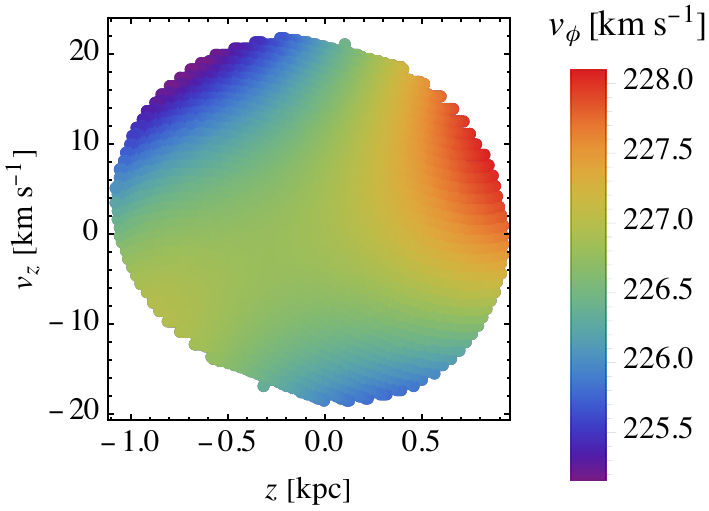} {\hspace{7pt}}&{\hspace{7pt}} \includegraphics[width=4cm]{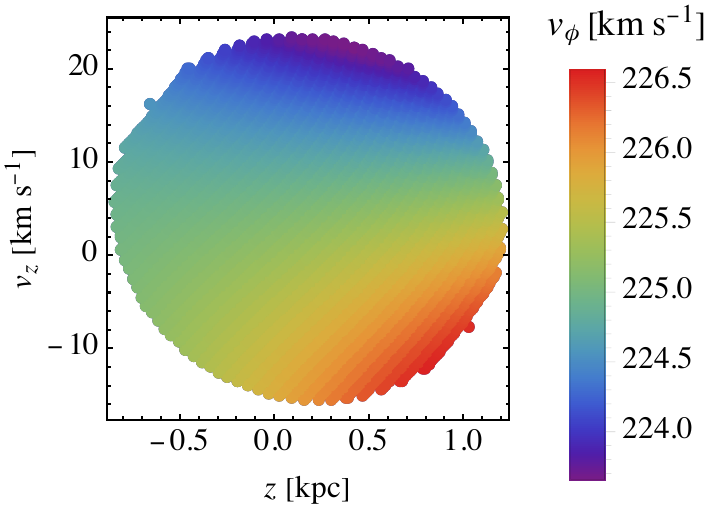} {\hspace{7pt}}&{\hspace{7pt}} \includegraphics[width=4cm]{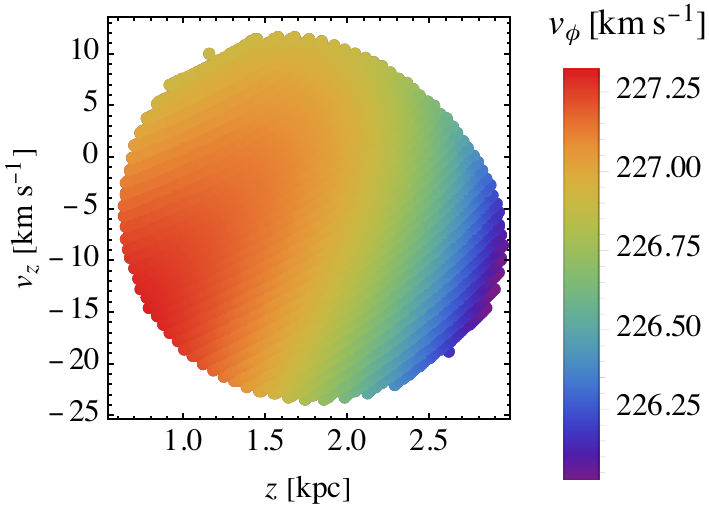} \\
    \end{tabular}
    \caption{Snapshots of the stellar phase space $(z, v_z)$ at galactocentric radius $R_0=9,12$ and 15\,kpc. Panels in the upper row show the unperturbed phase-space distribution, whereas the ones in the lower row display the phase space deformation in the presence of an oscillating dark matter halo.}
    \label{Fig3}
\end{figure*}

\subsection{Phase space signatures}

The results of the previous subsection provide the manifestations of the effects of the relative oscillations of the matter and dark matter halo in a model galaxy with parameters close to the Milky Way one. Fig.~\ref{Fig1} contains the plots of vertical perturbations of star orbits at different distances from the galactic center. Although this figure is very insightful, the long-term time evolution of star orbits cannot be observed directly. What is available in the catalogs of Gaia space observatory (see, e.g., \cite{Gaia3}) is a snapshot of a phase space of stars in the solar neighborhood. Therefore, it is necessary to model the effects of the dark matter halo oscillations for the coordinates of stars in the phase space. The phase space analysis appeared very fruitful for the studies of effects of incomplete phase mixing of the last massive merger by \cite{Belokurov-22} and \cite{Spiral}.

In this subsection, we consider the same model galaxy with the parameters described above, and numerically calculate trajectories of stars with the oscillation of dark matter halo being present (perturbation=on) and compare it with the similar solution with no such oscillation (perturbation=off). In the coordinate space, these trajectories are the solutions of Eq.~(\ref{NewtonEq}). We consider the  trajectories of stars at the distances $R_0\approx 9,12$ and 15\,kpc from GC understanding that the effect of DM halo oscillation grows with the distance from GC.

To produce the plots in Fig.~\ref{Fig3}, we have considered $N\approx 6200$ stars with approximately circular trajectories at the distance $R_0$ from GC and with the vertical displacement from the galactic plain ranging $|z_0| < 1\,\text{kpc}$. The $z$ components of the initial velocities of these stars vary as $|v_{z,0}| < 20\,\text{km\,s}^{-1}$. Then we allow these stars to evolve for the period $t = 1.6$\,Gyr which is approximately equal to four full periods of oscillations of dark matter halo (\ref{TNFW}), and take a snapshot of the phase space for all these trajectories. In Fig.~\ref{Fig3}, we present the colour map of the subspace $(z,v_z)$ of the full 6D phase space, with colour representing the azimuthal velocity component $v_\phi$. Comparison of the shapes and the colour patterns of the phase space snapshots with the perturbation on and off allows us to identify the signatures of the dark matter halo oscillations in the model galaxy. The subspace $(z,v_z)$ of the phase space is particularly interesting because the perturbation is applied in the $z$ direction in our model. 

To make the effect of the dark matter halo oscillation more apparent we artificially amplified the amplitude of oscillation (\ref{delta0}) 2.5 times. This allows us to observe the effect directly, without invoking statistical analysis. Note that the spread in $v_\phi$ is so small (a few km\,s$^{-1}$) because we considered trajectories of stars with vanishing initial spread in this component of the velocity, and the resulting spread is obtained as the result of the evolution. In reality, stars may have some initial distribution of $v_\phi$, but we ignore it for simplicity.

The left two plots in Fig.~\ref{Fig3} represent the phase space diagrams for trajectories of stars at the distance $R_0\approx 9$\,kpc from GC. Periods of these trajectories are not in the resonance with the DM halo oscillations (\ref{TNFW}), and the deformation of the prase space (bottom panel) is minor as compared with case with no perturbation (top panel). The colour pattern, however, reveals the asymmetry of the velocity component $v_\phi$. This asymmetry may be considered as a potential signature of the effect of oscillation of dark matter halo for stars with orbits out of the resonance with this oscillation. 

The right two plots in Fig.~\ref{Fig3} demonstrate the phase space diagrams for trajectories of stars at the distance $R_0\approx 15$\,kpc from GC. Periods of these trajectories are close to the period of DM halo oscillations (\ref{TNFW}). As a result, there is significant deformation of the phase space on the bottom panel as compared with the top one: the picture in the phase space is shifted towards the positive $z$ (the direction where the perturbation was initially applied) and negative $v_z$. Moreover, the colour pattern in this figure also has significant asymmetry when the DM halo oscillations are on. Both these effects may be considered as manifestations of the DM halo oscillation in the model galaxy. 

Comparing the two plots in the middle of Fig.~\ref{Fig3} we see that the bottom picture is shifted towards positive $z$ and $v_z$ and possesses apparent asymmetry in the color pattern. All these properties are common signatures of the dark matter halo oscillation.

It would be interesting to search for these effects in the actual data of the phase space snapshot provided by Gaia observatory \citep{Gaia3}. This task goes beyond the goals of the present letter and should be addressed elsewhere. Here we study this effect only theoretically in a model galaxy with the matter/dark matter composition close to the one of the Milky Way galaxy. It would be of interest to perform a realistic and accurate multi-body simulation of the dark matter halo oscillation and its manifestations on the star trajectories.

\section{Oscillations of the hot gas component}
\label{SecGas}

In the previous sections, we considered the galaxy model with two components represented by the matter and dark matter. More generally, one can distinguish also the hot ionized gas halo as an independent component. The mass density distributions of the matter and dark matter components may be taken in the from of the Plummer profile (\ref{Plummer}) and NFW profile (\ref{NFW}), respectively. Following \cite{HotGas1,HotGas2}, the density of the hot gas halo may be described by a spherically-symmetric $\beta$-profile,
\begin{equation}
    \rho_g(r) = \frac{\rho_0}{\left[1+ (r/r_c)^2 \right]^{3\beta/2}}\quad(0\leq r\leq R)\,.
    \label{gas}
\end{equation}
Typical values of the parameters in these model may be chosen as $\beta = 0.5$, $R=200$\,kpc and $r_c = 20\,\mbox{kpc}$. With $\rho_0 = 5\times 10^4\,M_\odot\mbox{kpc}^{-3}$, this density function is normalized to
\begin{equation}
    4\pi\int_0^R \rho_g(r)r^2 dr = M_g = 10^{11}\,M_\odot\,.
\end{equation}
Thus, this component is second in the mass in the galaxy after the dark matter component. It may play important role in the dynamics of the galaxy.

In equilibrium, each of the components creates its gravitational potential $\Phi_i(r)$, $i=m,d,g$ for matter, dark matter and gas halo, respectively. We will now assume that the matter and dark matter components are stationary while the gas component is displaced by a small vector $\boldsymbol{\delta}$ relative to the other two. Gas halo gravitational potential is perturbed as
\begin{equation}
    \Phi_g(|\boldsymbol{r}+\boldsymbol{\delta}|) = \Phi_g(r) + \boldsymbol{\delta}\cdot \nabla\Phi_g(r) +{\cal O}(\delta^2)\,,
\end{equation}
and the restoring force $\boldsymbol{F} = - k\boldsymbol{\delta}$ is specified by the spring constant analogous to (\ref{k-integral}):
\begin{equation}
    k = \frac{16\pi^2 G}{3} \int_0^R [\rho_m(r)+\rho_d(r)]\rho_g(r) r^2 dr\,.
\end{equation}
Substituting here the density distributions (\ref{Plummer}), (\ref{NFW}) and (\ref{gas}), we find the following value for the spring constant: $k \simeq 6.3 \times 10^{-7}\,M_\odot\mbox{yr}^{-2}$. The corresponding oscillation frequency $\omega_0 = \sqrt{k/\mu}$ and the period are
\begin{equation}
\omega_0\approx 2.6\,\mbox{Gyr}^{-1}\,,\qquad
T \approx 2.4 \, \mbox{Gyr}.
\label{omega-gas}
\end{equation}
This period is much larger than that in Eq.~(\ref{TNFW}) where the hot gas component has not been taken into account. 

The damping constant may be estimated with Eq.~(\ref{damping}), but $M_m$ should be replaced with $M_g$, and $\ln\Lambda\simeq1$:\footnote{The Coulomb logarithm takes into account the interaction range in the scattering process. Here we consider a small oscillation of the hot gas halo inside the dark matter component, which is different from the scattering process. Therefore, for this oscillation, it is likely that $\ln\Lambda\ll1$, and $\ln \Lambda\simeq1$ is rather an upper bound.}
\begin{equation}
    \gamma = 0.08\,.
\end{equation}
Thus, it is plausible that the gas component can do $\sim 10$ oscillations before the relaxation.

Consider now the gas halo with the center of mass coordinate in the galactic rest frame represented by a damped oscillation with angular velocity $\omega_0$ and amplitude $A\ll R_g$,
\begin{equation}
    \boldsymbol{r}_g(t) = \hat{\boldsymbol{z}} A e^{-\gamma\omega_0 t} \sin(\omega_0 t)\,.
\end{equation}
Stars in the galaxy move in the time-dependent gravitational potential
\begin{align}
    \Phi(\boldsymbol{r},t) = &\Phi_d(\boldsymbol{r}) + \Phi_m(\boldsymbol{r}) + \Phi_g(\boldsymbol{r}-\boldsymbol{r}_g(t)) \nonumber\\
    \approx& \Phi_d(\boldsymbol{r}) + \Phi_m(\boldsymbol{r}) + \Phi_g(\boldsymbol{r}) -\boldsymbol{r}_g(t) \cdot \nabla \Phi_g(\boldsymbol{r})\,.
\end{align}
Trajectories of stars in this time-dependent potential obey
\begin{equation}
    \ddot{\boldsymbol{r}} = - \nabla\Phi (\boldsymbol{r},t)\,.
\end{equation}
Considering $z$-component in this equation and linearizing it in $z$ and in $A$, we have a driven harmonic oscillation equation:
\begin{equation}
    \ddot z+ \omega^2 z = B e^{-\gamma\omega_0 t} \sin(\omega_0 t)\,,
\end{equation}
where
\begin{align}
    \omega^2 = & \partial_z^2 \Phi_m(\boldsymbol{R},0)+\partial_z^2 \Phi_d(\boldsymbol{R},0)+\partial_z^2 \Phi_g(\boldsymbol{R},0) \,,\\
    B =& A \partial_z^2 \Phi_g(\boldsymbol{R},0)\,,
\end{align}
with $\boldsymbol{R} = (x,y)$. 

Since $\omega=\omega(R)$, stars with orbits at some distance $R$ from GC could appear in resonance with the driving gas halo oscillations if 
\begin{equation}
    \omega(R) = \omega_0\,.
\end{equation}
For $\omega_0$ given by Eq.~(\ref{omega-gas}), this equation may be solved numerically, resulting $R\simeq 110$\,kpc. This distance is too far from the galactic disc. Therefore, unlike the oscillations of the galactic disc relative to the dark matter halo considered in Sec.~\ref{SecTrajectories}, the oscillation of the hot gas halo cannot cause significant perturbations of star orbits.


\section{Conclusions}

In this paper, we have estimated the effect of oscillation of matter in a galaxy relative to the dark matter halo. We have considered the case where such oscillations have a small amplitude as compared with the size of the galaxy. In this regime, the oscillations are approximately harmonic with a period comparable to the periods of star rotations in the galaxy. We argued that these oscillations are damped mainly through dynamical friction (the quality factor $\sim$ 10). It is plausible that such oscillations may be caused by collision of galaxies, although the question of possible origin of this effect requires further studies. The hypothesis of existence of these oscillations is confirmed by the analysis of \emph{Gaia} space observatory data by \cite{Belokurov-22} and numerical simulations in by \cite{Bland-Hawthorn_2025}. In this letter, we develop a simple analytical description of one possible contribution to such non-equilibrium dynamics.

In Sec.~\ref{SecTrajectories} we showed that oscillations of the dark matter halo relative to the matter component can produce significant perturbations to stellar orbits within the Galaxy. To illustrate this effect, we considered stars initially on circular orbits in the galactic plane ($z=0$) and examined the evolution of their vertical motion under the periodic forcing induced by the halo oscillation. We found that the vertical ($z$) components of these trajectories oscillate with a frequency that depends on the galactocentric radius. For stars at particular radii, this vertical frequency becomes commensurate with the driving frequency, leading to a resonance. The resulting amplification of vertical oscillations can, in principle, eject such stars from the galactic disk, potentially contributing to the population of runaway stars. Analogous oscillations of the hot gas halo component in galaxies should have significantly larger periods and are unlikely to produce significant perturbations of star orbits.

The anomalous vertical deviations of the stars in the disk of the Milky Way galaxy were observed by \cite{Poggio25}. Due to the specific shape of the deformation of the galactic disk, this effect was named ``The Great Wave.'' The absolute value of these deviations and the corresponding vertical component of the anomalous velocity were observed to be $|\Delta z|\lesssim 0.2$\,kpc and $|v_z|\lesssim 3$\,km\,s$^{-1}$. These values, as well as the qualitative wave-like morphology, are reproduced at the order-of-magnitude level in our model, see Fig.~\ref{Fig25}. The predicted velocity perturbation is smaller than the single-star intrinsic velocity scatter of the young Galactic disc, but may be accessible statistically as a coherent pattern in binned phase-space maps. This order-of-magnitude agreement suggests that an oscillating dark matter halo could contribute to Great-Wave-like vertical corrugations. It does not constitute a unique identification of the mechanism, since the observed Milky Way disc contains other non-axisymmetric perturbations with comparable or larger velocity amplitudes.

We have also performed a phase space analysis of the dark matter oscillations for star trajectories at the distances $R_0 = 9$\,kpc, 12\,kpc and 15\,kpc from GC. In the $(z,v_z)$ space, the dark matter oscillation manifests itself in the asymmetry of the phase diagrams as compared with the non-perturbed case, see Fig.~\ref{Fig3}. This asymmetry becomes especially apparent in the trajectories of stars with $R_0=15$\,kpc because of the resonance amplification of the perturbations. Searches of these effects in observational data like \cite{Gaia3} will be done elsewhere. Existence of the considered effect does not mean that it is easily  measurable, since it may be  hidden by a spread of stars velocities.

The simplified qualitative considerations presented in this paper may be generalized in the following ways:
\begin{enumerate}
    \item Cluster mergers provide clear examples in which collisionless and collisional matter components separate. In the Bullet Cluster \citep{Clowe:2003tk}, the collision is between two galaxy clusters, and the separation occurs between the dark-matter distribution and the hot intracluster gas. This environment is not directly analogous to the interstellar gas in a galactic disc, where hydrodynamical, feedback, and star-formation processes are much more complex. Nevertheless, such systems illustrate that gravitationally bound collisionless and collisional components need not remain exactly co-centered after a strong perturbation. It would be interesting to study whether analogous, but galaxy-scale, perturbations of gaseous haloes could induce misalignments between the stellar, cold-gas and dark-matter components considered in this paper.
    \item It is tempting to perform numeric simulations for the trajectories of stars in the Milky Way galaxy assuming that the dark matter halo is non-static but oscillating near the center of mass. We expect that such oscillations may cause deformations of the galactic disc similar to the one observed by \cite{Poggio25}. Such oscillations, if exist, should manifest themselves in anomalies of velocities of stars in the galaxy, such as the density waves  and runaway stars which have orbit period in resonance with oscillations.
    \item Damping of oscillations may be compensated by new collisions exciting oscillations again. In the harmonic regime, the frequency of the oscillations does not depend on the strength of each kick, so the resonance condition between the star period of rotation and oscillation frequency is not violated.
    \item We have assumed a single smooth dark-matter halo. More complicated halo phase-space structure may arise after mergers or satellite passages. Cluster mergers such as the Bullet Cluster demonstrate the separation of collisionless and collisional components on cluster scales, but they should not be described as galaxy--galaxy collisions and are not a direct analogue of the baryonic gas dynamics inside galactic discs.
\end{enumerate}
We leave all these question for future studies.

\section*{Acknowledgements}
We are grateful to Sergei Popov and Vasily Belokurov for useful discussions and valuable comments. This work was supported by the Australian Research Council Grants No.\ DP230101058.

\bibliographystyle{mnras}
\bibliography{literature}

@article{Clowe:2006eq,
    author = "Clowe, Douglas and Bradac, Marusa and Gonzalez, Anthony H. and Markevitch, Maxim and Randall, Scott W. and Jones, Christine and Zaritsky, Dennis",
    title = "{A direct empirical proof of the existence of dark matter}",
    eprint = "astro-ph/0608407",
    archivePrefix = "arXiv",
    reportNumber = "SLAC-PUB-12078",
    journal = "Astrophys. J. Lett.",
    volume = "648",
    pages = "L109--L113",
    year = "2006",
    doi = {10.1086/508162}
}

@article{Poggio25,
	author = {Poggio, E. and Khanna, S. and Drimmel, R. and Zari, E. and D’Onghia, E. and Lattanzi, M. G. and Palicio, P. A. and Recio-Blanco, A. and Thulasidharan, L.},
	title = {The great wave - Evidence of a large-scale vertical corrugation propagating outwards in the Galactic disc},
	url= "https://doi.org/10.1051/0004-6361/202451668",
	journal = {A\&A},
	year = 2025,
	volume = 699,
	pages = "A199",
    eprint = "2407.18659",
    archivePrefix = "arXiv",
    primaryClass = "astro-ph.GA",
    doi = "10.1051/0004-6361/202451668"
}

@article{Kuhlen2013,
    author = "Kuhlen, Michael and Guedes, Javiera and Pillepich, Annalisa and Madau, Piero and Mayer, Lucio",
    title = "{An off-center density peak in the Milky Way's dark matter halo?}",
    eprint = "1208.4844",
    archivePrefix = "arXiv",
    primaryClass = "astro-ph.GA",
    journal = "Astrophys. J.",
    volume = "765",
    pages = "10",
    year = "2013",
    doi = "10.1088/0004-637X/765/1/10"
}

@article{Schaller2015,
    author = "Schaller, Matthieu and Robertson, Andrew and Massey, Richard and Bower, Richard G. and Eke, Vincent R.",
    title = "{The offsets between galaxies and their dark matter in $\Lambda$ cold dark matter}",
    eprint = "1505.05470",
    archivePrefix = "arXiv",
    primaryClass = "astro-ph.CO",
    journal = "Mon. Not. Roy. Astron. Soc.",
    volume = "453",
    number = "1",
    pages = "L58--L62",
    year = "2015",
    doi = "10.1093/mnrasl/slv104"
}

@article{Massey2015,
    author = "Massey, Richard and others",
    title = "{The behaviour of dark matter associated with four bright cluster galaxies in the 10 kpc core of Abell 3827}",
    eprint = "1504.03388",
    archivePrefix = "arXiv",
    primaryClass = "astro-ph.CO",
    journal = "Mon. Not. Roy. Astron. Soc.",
    volume = "449",
    number = "4",
    pages = "3393--3406",
    year = "2015",
    doi = "10.1093/mnras/stv467"
}

@other{Su:2012ft,
    author = "Su, Meng and Finkbeiner, Douglas P.",
    title = "Strong evidence for gamma-ray line emission from the Inner Galaxy",
    eprint = "1206.1616",
    archivePrefix = "arXiv",
    primaryClass = "astro-ph.HE",
    month = "6",
    year = "2012"
}

@article{Prasad2017,
       author = {Prasad, C. and Jog, C. J.},
        title = "{Off-center dark matter halo leading to strong central disk lopsidedness}",
      journal = {A\&A},
         year = 2017,
        month = apr,
       volume = {600},
        pages = {A17},
archivePrefix = {arXiv},
       eprint = {1610.01702},
 primaryClass = {astro-ph.GA},
    doi = {10.1051/0004-6361/201630071}
}

@article{Clowe:2003tk,
    author = "Clowe, Douglas and Gonzalez, Anthony and Markevitch, Maxim",
    title = "{Weak lensing mass reconstruction of the interacting cluster 1E0657-558: Direct evidence for the existence of dark matter}",
    eprint = "astro-ph/0312273",
    archivePrefix = "arXiv",
    journal = "Astrophys. J.",
    volume = "604",
    pages = "596--603",
    year = "2004",
    doi = "10.1086/381970"
}

@ARTICLE{Chandrasekhar43,
       author = {Chandrasekhar, S.},
        title = "{Dynamical friction. I. General considerations: the coefficient of dynamical friction}",
      journal = {Astrophys. J.},
         year = 1943,
        month = mar,
       volume = {97},
        pages = {255},
          doi = {10.1086/144517}
}

@article{Bland-Hawthorn_2025,
url = {https://doi.org/10.3847/1538-4357/ae0931},
year = {2025},
month = {nov},
publisher = {The American Astronomical Society},
volume = {994},
number = {1},
pages = {22},
author = {Bland-Hawthorn, Joss and Tepper-Garcia, Thor and Agertz, Oscar and Federrath, Christoph and Haywood, Misha and di Matteo, Paola and Bedding, Timothy R. and Tsukui, Takafumi and Wisnioski, Emily and Ness, Melissa and Freeman, Ken},
title = {Turbulent gas-rich disks at high redshift: Origin of thick stellar disks through 3D “Baryon Sloshing”},
journal = {Astrophys. J.},
doi = {10.3847/1538-4357/ae0931},
    eprint = "2502.01895",
    archivePrefix = "arXiv",
    primaryClass = "astro-ph.GA"
}

@article{Vasiliev_2022,
url = {https://doi.org/10.3847/1538-4357/ac4fbc},
year = {2022},
month = {feb},
publisher = {The American Astronomical Society},
volume = {926},
number = {2},
pages = {203},
author = {Vasiliev, Eugene and Belokurov, Vasily and Evans, N. Wyn},
title = {Radialization of satellite orbits in galaxy mergers},
journal = {Astrophys. J.},
    eprint = "2108.00010",
    archivePrefix = "arXiv",
    primaryClass = "astro-ph.GA",
    doi = "10.3847/1538-4357/ac4fbc"
}

@article{Belokurov-22,
    author = {Belokurov, Vasily and Vasiliev, Eugene and Deason, Alis J and Koposov, Sergey E and Fattahi, Azadeh and Dillamore, Adam M and Davies, Elliot Y and Grand, Robert J J},
    title = {Energy wrinkles and phase-space folds of the last major merger},
    journal = {Mon. Not. Roy. Astron. Soc.},
    volume = {518},
    number = {4},
    pages = {6200-6215},
    year = {2022},
    month = {11},
    issn = {0035-8711},
    doi = {10.1093/mnras/stac3436},
    url = {https://doi.org/10.1093/mnras/stac3436},
    eprint = "2208.11135",
    archivePrefix = "arXiv",
    primaryClass = "astro-ph.GA"
}

@article{HotGas1,
       author = {{Cavaliere}, A. and {Fusco-Femiano}, R.},
        title = "{X-rays from hot plasma in clusters of galaxies.}",
      journal = {A\&A},
     year = 1976,
        month = may,
       volume = {49},
        pages = {137-144}
}

@article{HotGas2,
       author = {{Cavaliere}, A. and {Fusco-Femiano}, R.},
        title = "{The distribution of hot gas in clusters of galaxies}",
      journal = {A\&A},
         year = 1978,
        month = nov,
       volume = {70},
        pages = {677}
}

@article{Gaia3,
	author = {{Gaia Collaboration} and {Drimmel, R.} and {Romero-G\'omez, M.} and {Chemin, L.} and {Ramos, P.} and {Poggio, E.} and {Ripepi, V.} and {Andrae, R.} and {Blomme, R.} and {Cantat-Gaudin, T.} and {Castro-Ginard, A.} and {Clementini, G.} and {Figueras, F.} and {Fouesneau, M.} and {Fr\'emat, Y.} and {Jardine, K.} and {Khanna, S.} and {Lobel, A.} and {Marshall, D. J.} and {Muraveva, T.} and {Brown, A. G. A.} and {Vallenari, A.} and {Prusti, T.} and {de Bruijne, J. H. J.} and {Arenou, F.} and {Babusiaux, C.} and {Biermann, M.} and {Creevey, O. L.} and {Ducourant, C.} and {Evans, D. W.} and {Eyer, L.} and {Guerra, R.} and {Hutton, A.} and {Jordi, C.} and {Klioner, S. A.} and {Lammers, U. L.} and {Lindegren, L.} and {Luri, X.} and {Mignard, F.} and {Panem, C.} and {Pourbaix, D.} and {Randich, S.} and {Sartoretti, P.} and {Soubiran, C.} and {Tanga, P.} and {Walton, N. A.} and {Bailer-Jones, C. A. L.} and {Bastian, U.} and {Jansen, F.} and {Katz, D.} and {Lattanzi, M. G.} and {van Leeuwen, F.} and {Bakker, J.} and {Cacciari, C.} and {Casta\~neda, J.} and {De Angeli, F.} and {Fabricius, C.} and {Galluccio, L.} and {Guerrier, A.} and {Heiter, U.} and {Masana, E.} and {Messineo, R.} and {Mowlavi, N.} and {Nicolas, C.} and {Nienartowicz, K.} and {Pailler, F.} and {Panuzzo, P.} and {Riclet, F.} and {Roux, W.} and {Seabroke, G. M.} and {Sordo, R.} and {Th\'evenin, F.} and {Gracia-Abril, G.} and {Portell, J.} and {Teyssier, D.} and {Altmann, M.} and {Audard, M.} and {Bellas-Velidis, I.} and {Benson, K.} and {Berthier, J.} and {Burgess, P. W.} and {Busonero, D.} and {Busso, G.} and {C\'anovas, H.} and {Carry, B.} and {Cellino, A.} and {Cheek, N.} and {Damerdji, Y.} and {Davidson, M.} and {de Teodoro, P.} and {Nu\~nez Campos, M.} and {Delchambre, L.} and {Dell\'{}Oro, A.} and {Esquej, P.} and {Fern\'andez-Hern\'andez, J.} and {Fraile, E.} and {Garabato, D.} and {Garc\'{\i}a-Lario, P.} and {Gosset, E.} and {Haigron, R.} and {Halbwachs, J.-L.} and {Hambly, N. C.} and {Harrison, D. L.} and {Hern\'andez, J.} and {Hestroffer, D.} and {Hodgkin, S. T.} and {Holl, B.} and {Jan\ss{}en, K.} and {Jevardat de Fombelle, G.} and {Jordan, S.} and {Krone-Martins, A.} and {Lanzafame, A. C.} and {L\"offler, W.} and {Marchal, O.} and {Marrese, P. M.} and {Moitinho, A.} and {Muinonen, K.} and {Osborne, P.} and {Pancino, E.} and {Pauwels, T.} and {Recio-Blanco, A.} and {Reyl\'e, C.} and {Riello, M.} and {Rimoldini, L.} and {Roegiers, T.} and {Rybizki, J.} and {Sarro, L. M.} and {Siopis, C.} and {Smith, M.} and {Sozzetti, A.} and {Utrilla, E.} and {van Leeuwen, M.} and {Abbas, U.} and {\'Abrah\'am, P.} and {Abreu Aramburu, A.} and {Aerts, C.} and {Aguado, J. J.} and {Ajaj, M.} and {Aldea-Montero, F.} and {Altavilla, G.} and {\'Alvarez, M. A.} and {Alves, J.} and {Anders, F.} and {Anderson, R. I.} and {Anglada Varela, E.} and {Antoja, T.} and {Baines, D.} and {Baker, S. G.} and {Balaguer-N\'u\~nez, L.} and {Balbinot, E.} and {Balog, Z.} and {Barache, C.} and {Barbato, D.} and {Barros, M.} and {Barstow, M. A.} and {Bartolom\'e, S.} and {Bassilana, J.-L.} and {Bauchet, N.} and {Becciani, U.} and {Bellazzini, M.} and {Berihuete, A.} and {Bernet, M.} and {Bertone, S.} and {Bianchi, L.} and {Binnenfeld, A.} and {Blanco-Cuaresma, S.} and {Boch, T.} and {Bombrun, A.} and {Bossini, D.} and {Bouquillon, S.} and {Bragaglia, A.} and {Bramante, L.} and {Breedt, E.} and {Bressan, A.} and {Brouillet, N.} and {Brugaletta, E.} and {Bucciarelli, B.} and {Burlacu, A.} and {Butkevich, A. G.} and {Buzzi, R.} and {Caffau, E.} and {Cancelliere, R.} and {Carballo, R.} and {Carlucci, T.} and {Carnerero, M. I.} and {Carrasco, J. M.} and {Casamiquela, L.} and {Castellani, M.} and {Chaoul, L.} and {Charlot, P.} and {Chiaramida, V.} and {Chiavassa, A.} and {Chornay, N.} and {Comoretto, G.} and {Contursi, G.} and {Cooper, W. J.} and {Cornez, T.} and {Cowell, S.} and {Crifo, F.} and {Cropper, M.} and {Crosta, M.} and {Crowley, C.} and {Dafonte, C.} and {Dapergolas, A.} and {David, P.} and {de Laverny, P.} and {De Luise, F.} and {De March, R.} and {De Ridder, J.} and {de Souza, R.} and {de Torres, A.} and {del Peloso, E. F.} and {del Pozo, E.} and {Delbo, M.} and {Delgado, A.} and {Delisle, J.-B.} and {Demouchy, C.} and {Dharmawardena, T. E.} and {Di Matteo, P.} and {Diakite, S.} and {Diener, C.} and {Distefano, E.} and {Dolding, C.} and {Enke, H.} and {Fabre, C.} and {Fabrizio, M.} and {Faigler, S.} and {Fedorets, G.} and {Fernique, P.} and {Fournier, Y.} and {Fouron, C.} and {Fragkoudi, F.} and {Gai, M.} and {Garcia-Gutierrez, A.} and {Garcia-Reinaldos, M.} and {Garc\'{\i}a-Torres, M.} and {Garofalo, A.} and {Gavel, A.} and {Gavras, P.} and {Gerlach, E.} and {Geyer, R.} and {Giacobbe, P.} and {Gilmore, G.} and {Girona, S.} and {Giuffrida, G.} and {Gomel, R.} and {Gomez, A.} and {Gonz\'alez-N\'u\~nez, J.} and {Gonz\'alez-Santamar\'{\i}a, I.} and {Gonz\'alez-Vidal, J. J.} and {Granvik, M.} and {Guillout, P.} and {Guiraud, J.} and {Guti\'errez-S\'anchez, R.} and {Guy, L. P.} and {Hatzidimitriou, D.} and {Hauser, M.} and {Haywood, M.} and {Helmer, A.} and {Helmi, A.} and {Sarmiento, M. H.} and {Hidalgo, S. L.} and {Hladczuk, N.} and {Hobbs, D.} and {Holland, G.} and {Huckle, H. E.} and {Jasniewicz, G.} and {Jean-Antoine Piccolo, A.} and {Jim\'enez-Arranz, \'O.} and {Juaristi Campillo, J.} and {Julbe, F.} and {Karbevska, L.} and {Kervella, P.} and {Kordopatis, G.} and {Korn, A. J.} and {K\'osp\'al, \'A} and {Kostrzewa-Rutkowska, Z.} and {Kruszy\'{}nska, K.} and {Kun, M.} and {Laizeau, P.} and {Lambert, S.} and {Lanza, A. F.} and {Lasne, Y.} and {Le Campion, J.-F.} and {Lebreton, Y.} and {Lebzelter, T.} and {Leccia, S.} and {Leclerc, N.} and {Lecoeur-Taibi, I.} and {Liao, S.} and {Licata, E. L.} and {Lindstr\o{}m, H. E. P.} and {Lister, T. A.} and {Livanou, E.} and {Lorca, A.} and {Loup, C.} and {Madrero Pardo, P.} and {Magdaleno Romeo, A.} and {Managau, S.} and {Mann, R. G.} and {Manteiga, M.} and {Marchant, J. M.} and {Marconi, M.} and {Marcos, J.} and {Marcos Santos, M. M. S.} and {Mar\'{\i}n Pina, D.} and {Marinoni, S.} and {Marocco, F.} and {Martin Polo, L.} and {Mart\'{\i}n-Fleitas, J. M.} and {Marton, G.} and {Mary, N.} and {Masip, A.} and {Massari, D.} and {Mastrobuono-Battisti, A.} and {Mazeh, T.} and {McMillan, P. J.} and {Messina, S.} and {Michalik, D.} and {Millar, N. R.} and {Mints, A.} and {Molina, D.} and {Molinaro, R.} and {Moln\'ar, L.} and {Monari, G.} and {Mongui\'o, M.} and {Montegriffo, P.} and {Montero, A.} and {Mor, R.} and {Mora, A.} and {Morbidelli, R.} and {Morel, T.} and {Morris, D.} and {Murphy, C. P.} and {Musella, I.} and {Nagy, Z.} and {Noval, L.} and {Oca\~na, F.} and {Ogden, A.} and {Ordenovic, C.} and {Osinde, J. O.} and {Pagani, C.} and {Pagano, I.} and {Palaversa, L.} and {Palicio, P. A.} and {Pallas-Quintela, L.} and {Panahi, A.} and {Payne-Wardenaar, S.} and {Pe\~nalosa Esteller, X.} and {Penttil\"a, A.} and {Pichon, B.} and {Piersimoni, A. M.} and {Pineau, F.-X.} and {Plachy, E.} and {Plum, G.} and {Prsa, A.} and {Pulone, L.} and {Racero, E.} and {Ragaini, S.} and {Rainer, M.} and {Raiteri, C. M.} and {Ramos-Lerate, M.} and {Re Fiorentin, P.} and {Regibo, S.} and {Richards, P. J.} and {Rios Diaz, C.} and {Riva, A.} and {Rix, H.-W.} and {Rixon, G.} and {Robichon, N.} and {Robin, A. C.} and {Robin, C.} and {Roelens, M.} and {Rogues, H. R. O.} and {Rohrbasser, L.} and {Rowell, N.} and {Royer, F.} and {Ruz Mieres, D.} and {Rybicki, K. A.} and {Sadowski, G.} and {S\'aez N\'u\~nez, A.} and {Sagrist\`a Sell\'es, A.} and {Sahlmann, J.} and {Salguero, E.} and {Samaras, N.} and {Sanchez Gimenez, V.} and {Sanna, N.} and {Santove\~na, R.} and {Sarasso, M.} and {Schultheis, M. S.} and {Sciacca, E.} and {Segol, M.} and {Segovia, J. C.} and {S\'egransan, D.} and {Semeux, D.} and {Shahaf, S.} and {Siddiqui, H. I.} and {Siebert, A.} and {Siltala, L.} and {Silvelo, A.} and {Slezak, E.} and {Slezak, I.} and {Smart, R. L.} and {Snaith, O. N.} and {Solano, E.} and {Solitro, F.} and {Souami, D.} and {Souchay, J.} and {Spagna, A.} and {Spina, L.} and {Spoto, F.} and {Steele, I. A.} and {Steidelm\"uller, H.} and {Stephenson, C. A.} and {S\"uveges, M.} and {Surdej, J.} and {Szabados, L.} and {Szegedi-Elek, E.} and {Taris, F.} and {Taylor, M. B.} and {Teixeira, R.} and {Tolomei, L.} and {Tonello, N.} and {Torra, F.} and {Torra, J.} and {Torralba Elipe, G.} and {Trabucchi, M.} and {Tsounis, A. T.} and {Turon, C.} and {Ulla, A.} and {Unger, N.} and {Vaillant, M. V.} and {van Dillen, E.} and {van Reeven, W.} and {Vanel, O.} and {Vecchiato, A.} and {Viala, Y.} and {Vicente, D.} and {Voutsinas, S.} and {Weiler, M.} and {Wevers, T.} and {Wyrzykowski, L.} and {Yoldas, A.} and {Yvard, P.} and {Zhao, H.} and {Zorec, J.} and {Zucker, S.} and {Zwitter, T.}},
	title = {Gaia Data Release 3 - Mapping the asymmetric disc of the Milky Way},
	DOI= "10.1051/0004-6361/202243797",
	url= "https://doi.org/10.1051/0004-6361/202243797",
	journal = {A\&A},
	year = 2023,
	volume = 674,
	pages = "A37",
}

@article{Spiral,
	author = {Antoja, T. and Ramos, P. and Garc\'{\i}a-Conde, B. and Bernet, M. and Laporte, C. F. P. and Katz, D.},
	title = {The phase spiral in Gaia DR3},
	DOI= "10.1051/0004-6361/202245518",
	url= "https://doi.org/10.1051/0004-6361/202245518",
	journal = {A\&A},
	year = 2023,
	volume = 673,
	pages = "A115",
}

@article{Jee_2007,
doi = {10.1086/517498},
url = {https://doi.org/10.1086/517498},
year = {2007},
month = {jun},
publisher = {},
volume = {661},
number = {2},
pages = {728},
author = {Jee, M. J. and Ford, H. C. and Illingworth, G. D. and White, R. L. and Broadhurst, T. J. and Coe, D. A. and Meurer, G. R. and van der Wel, A. and Benítez, N. and Blakeslee, J. P. and Bouwens, R. J. and Bradley, L. D. and Demarco, R. and Homeier, N. L. and Martel, A. R. and Mei, S.},
title = {Discovery of a Ringlike Dark Matter Structure in the Core of the Galaxy Cluster Cl 0024+17},
journal = {The Astrophysical Journal}
}

@article{Jee_2010,
doi = {10.1088/0004-637X/717/1/420},
url = {https://doi.org/10.1088/0004-637X/717/1/420},
year = {2010},
month = {jun},
publisher = {The American Astronomical Society},
volume = {717},
number = {1},
pages = {420},
author = {Jee, M. J.},
title = {TRACING THE PECULIAR DARK MATTER STRUCTURE IN THE GALAXY CLUSTER Cl 0024+17 WITH INTRACLUSTER STARS AND GAS},
journal = {The Astrophysical Journal}
}

\bsp
\label{lastpage}

\end{document}